\documentclass[english]{article}

\usepackage{array}
\usepackage{amsmath}
\usepackage{bm}
\usepackage{cancel}
\usepackage{graphicx}

\usepackage{color}
\definecolor{headingcolour}{RGB}{50, 150, 190}

\usepackage{tikz}
\usetikzlibrary{decorations.pathmorphing}
\usetikzlibrary{patterns}

\usepackage{babel}

\usepackage[authoryear]{natbib}

\usepackage{setspace}

\usepackage{quattrocento}

\usepackage{sectsty}
\allsectionsfont{\fontfamily{phv}\selectfont}

\usepackage[T1]{fontenc} 

\usepackage[utf8]{inputenc}
\usepackage[bottom]{footmisc}

\usepackage[margin=1in]{geometry}
\usepackage{caption}
\DeclareCaptionFormat{label}{\textbf{#1 #2} #3\hrulefill}
\usepackage[compact]{titlesec}
\titlespacing*{\section}{0pt}{0pt}{*0}
\titlespacing*{\subsection}{0pt}{0pt}{*0}

\usepackage{graphicx}

\usepackage{pifont}

\usepackage{enumitem}
\setlist{nosep}

\renewenvironment{abstract}{%
	\hspace{0.025\linewidth}\begin{minipage}{0.95\textwidth}
		\rule{\textwidth}{1pt}\small\selectfont}
	{\vspace{-0.5em}\par\noindent\rule{\textwidth}{1pt}\end{minipage}\vspace{1em}}
\makeatletter

\renewcommand{\maketitle}{\bgroup\setlength{\parindent}{0pt}
	\thispagestyle{empty}
	\begin{flushleft}
		{\bf \fontfamily{phv}\selectfont \LARGE \@title}
		
		\bf \fontfamily{phv}\selectfont \@author
	\end{flushleft}\egroup
}
\makeatother
\usepackage{siunitx}
\usepackage[%
    colorlinks=true,
    pdfborder={0 0 0},
    allcolors =cyan
]{hyperref}
\usepackage{dingbat}

\title{Travel time and energy dissipation minima in heterogeneous subsurface flows}

\author{Scott K. Hansen\footnote{Zuckerberg Institute for Water Research, Ben-Gurion University of the Negev}, Daniel O'Malley\footnote{EES-16, Los Alamos National Laboratory}}

\begin{document}
\doublespacing
\maketitle
\onehalfspacing

\begin{abstract}
	We establish a number of results concerning conditions for minimum energy dissipation and advective travel time in porous and fractured media. First, we establish a pair of converse results concerning fluid motion along a streamline between two points of fixed head: the minimal advective time is achieved under conditions of constant energy dissipation, and minimal energy dissipation is achieved under conditions of constant velocity along the streamline (implying homogeneous conductivity in the vicinity of the streamline). We also show directly by means of variational methods that minimum advection time along a streamline with a given average conductivity is achieved when the conductivity is constant. Finally, we turn our attention to minimum advection time and energy dissipation in parallel and sequential fracture systems governed by the cubic law: for which fracture cross-section and conductivity are intimately linked. We show that, as in porous domains, flow partitioning between different pathways always acts to minimize system energy dissipation. Finally, we consider minimum advection time as a function of aperture distribution in a sequence of fracture segments. We show that, for a fixed average aperture, a uniform-aperture system displays the shortest advection time. However, we also show that any sufficiently small small perturbations in aperture away from uniformity always act to reduce advection time.
\end{abstract}

\section{Introduction}
In the mid-1800s Henry Darcy formulated the empirical law bearing his name: namely that groundwater flux scales linearly with hydraulic head drop and a factor that represents a porous medium's intrinsic resistance to flow \citep{Brown2002}. On a one-dimensional manifold, the law has the simple form
\begin{equation}
	q(x) = -k(x)h'(x),
\end{equation}
where $q\ \mathrm{[LT^{-1}]}$ is volumetric flux, $k\ \mathrm{[LT^{-1}]}$ is hydraulic conductivity, and $h'$ represents the first spatial derivative of hydraulic head, $h\ \mathrm{[L]}$. Subsequently, \cite{Hubbert1940} reformulated Darcy's empirical equation in terms of potential flow, making explicit the relationships between various components of hydraulic head and energy density. In this conception, hydraulic head---which in Darcy's theory was an empirical measurement of water level in a piezometer tube---is understood as proportional to energy per unit volume, with constant of proportionality $\frac{1}{\rho g}$. Here, $\rho\ \mathrm{[ML^{-3}]}$ is mass density of fluid and $g\ \mathrm{[LT^{-2}]}$ is acceleration due to gravity. For example: gravitational potential energy density is $\rho g z$, whereas elevation head is $h_z \equiv z$, defined as elevation above datum. Similarly, kinetic energy density is $\frac{\rho v^2}{2}$, where $v$ is fluid velocity, versus velocity head $h_v \equiv \frac{v^2}{2g}$. Pressure, $P$, is potential energy density, versus pressure head $h_p \equiv \frac{P}{\rho g}$. Total head at location $x$ is $h(x) = h_z(x) + h_p(x) + h_v(x)$.

Viewed this way, groundwater flow dissipates energy, as porous media does work upon it resisting its movement. It is then natural to conceive of groundwater flow as configuring itself to minimize energy dissipation subject to boundary conditions. \cite{Narasimhan2008} explicitly states this as a general postulate and proved by variational techniques that the steady-state groundwater flow equation minimizes energy dissipation \citep{Narasimhan1999}. This principle was also asserted by \cite{Hergarten2014}, who employed it to study groundwater flow patterns, assuming intrinsic parameters at aquifer scale were configured in an approximately energy-minimizing way. Concretely, we write the energy dissipation functional
\begin{equation}
	E=\int_{x_1}^{x_2}k(x)h'^2(x)dx,
	\label{eq: E}
\end{equation}
where $E\ \mathrm{[L^2T^{-1}]}$ is proportional to energy dissipated along a stream tube flowing from $x_1$ to $x_2$. The same principle, in which energy dissipation is equal to the product of flow rate and head change has also been employed in the design of optimal pumping regimes \citep[e.g.,][]{Ahlfeld2015}

Conditions for minimum advective travel time have apparently received less attention. An exception is the work of \cite{Kacimov2009}, in which the authors establish lower bounds on advection times between two points in heterogeneous porous media by use of the Cauchy-Schwartz inequality. Minimum travel time through fracture networks has also been studied by a number of authors. \cite{Sweeney2023} note the importance of minimum travel time in understanding flow and transport processes in fractured systems, highlighting that transport can be slowed due to the interplay between fracture connectivity and aperture variability. \cite{Viswanathan2018} show that identifying shortest paths in the network can significantly reduce computational costs while accurately estimating travel times. 

While travel time is important, many works such as by \cite{Berkowitz1998}, \cite{Frampton2011}, and \cite{Edery2016} show that other aspects of transport such as tailing behavior are also important. Still, minimum travel times are key to applications such as contaminant transport, because they can put bounds on, e.g., how soon a contaminant source upstream can impact water supplies downstream, and are relevant to groundwater age dynamics \citep[e.g.,][]{Varni1998}.

In the following, we continue investigations into minimum energy dissipation and travel time in the context of optimal groundwater flow patterns in porous and fractured media. Our analyses are essentially in one spatial dimension, along streamlines of the Darcy flow or fracture paths, but we develop conclusions that are relevant in higher dimensions.

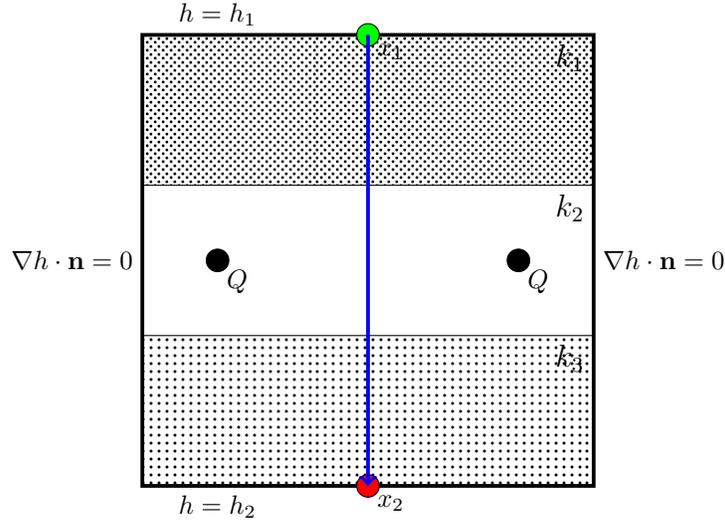
\begin{figure}
		\centering
	\begin{tikzpicture}
		\draw[line width=0.5mm] (0,0) rectangle (6,6);
		\draw[pattern=crosshatch dots] (0,4) rectangle (6,6);
		\node at (6,6)[below left]{\large $k_1$};
		\node at (6,4)[below left]{\large $k_2$};
		\draw[pattern=dots] (0,0) rectangle (6,2);
		\node at (6,2)[below left]{\large $k_3$};
		\draw [fill=black] (1,3) circle [radius=1.5mm] node [below right] {$Q$};
		\draw [fill=black] (5,3) circle [radius=1.5mm] node [below right] {$Q$};
		\draw [fill=green] (3,6) circle [radius=1.5mm] node [below right] {$x_1$};
		\draw [fill=red] (3,0) circle [radius=1.5mm] node [below right] {$x_2$};
		\draw[->, line width=0.5mm, blue] (3,6) -- (3,0);
		\node at (1,6) [above] {$h=h_1$};
		\node at (1,0) [below] {$h=h_2$};
		\node at (6,3) [right] {$\nabla h \cdot \mathbf{n}=0$};
		\node at (0,3) [left] {$\nabla h \cdot \mathbf{n}=0$};
	\end{tikzpicture}
	\caption{Toy system model of flow though zonated heterogeneous media, featuring three different conductivities, $k_1$, $k_2$, and $k_3$. Regardless of pumping rate, $Q$, at the two wells, the streamline from $x_1$ to  $x_2$ remains unchanged. However, the head distribution along the streamline is influenced by $Q$.}
	\label{fig: toy}
\end{figure}

In Section \ref{sec: cv mtt}, we consider the head distribution along a streamline that minimizes advective travel time. In Section \ref{sec: hd med}, we consider the head distribution that minimizes energy dissipation along a streamline. In Section \ref{sec: k mtt} we present a variational argument why a homogeneous hydraulic conductivity distribution minimizes travel time under natural gradient. In Section \ref{sec: md fracture}, we show that where multiple flow paths exist between two points in a fracture network that flow partitions itself to minimize energy dissipation. In Section \ref{sec: cubic}, we consider the flow in sequential fracture segments. We show that sufficiently small local perturbations from homogeneous aperture always reduce travel time, but also that in a series of fracture segments with a \textit{fixed} mean aperture and head drop, the greatest flow rate is achieved when all the segments have the same aperture.

\section{Head distribution for minimum travel time in porous media}
\label{sec: cv mtt}
\subsection{Formulation}
	We consider flow along a fixed streamline of the Darcy flow through a heterogeneous conductivity field in any number of dimensions, asking what head distribution along that streamline minimizes the advective travel time along it as a function of the hydraulic conductivities it experiences. For concreteness, Figure \ref{fig: toy} illustrates a toy system in which pumping affects head distribution along a fixed streamline. While general perturbations in head distribution may alter streamline paths in two and three dimensions, and thus the conductivities experienced, this remains an interesting investigation in arbitrary dimensions for two reasons. First, the analysis leads us to a general optimality condition relating energy dissipation and travel rates, valid in any number of dimensions. Second, it is well established that even under minor heterogeneity, profound flow channeling manifests \citep{Edery2014}, and that the overall flow pattern is determined by the interaction of the conductivity field and the large-scale head gradient \citep{Pozdniakov2004}. Thus, it is likely reasonable to treat higher dimensional flows as following quasi-fixed 1D paths for the purpose of manipulating local head gradients to enhance convection rates.

	Formally, we aim to minimize the functional
	\begin{equation}
	T=-\int_{x_1}^{x_2}\frac{\theta(x) dx}{k(x)h'(x)},
	\label{eq: T}
	\end{equation}
	representing the advection time $\mathrm{[T]}$ to between points $x_1$ and $x_2$, by optimal selection of $h(x)$. Here, $\theta \mathrm{[-]}$ represents porosity and $h(x)$ is assumed to have known $h(x_1)$ and $h(x_2)$ and to be be strictly decreasing from $x_1$ to $x_2$, but otherwise arbitrary.
	
\subsection{Minimum travel time is achieved under conditions of constant dissipation}	
	
	Euler considered selection of $h$ to minimize an arbitrary functional
	\begin{equation}
		T=\int_{x_1}^{x_2} F(x,h,h') dx.
	\end{equation}
	The approach is based on expanding $h = h_0 + \epsilon\eta$, where $h_0$ is the unknown optimal solution, $\epsilon$ is a small scalar, and $\eta$ is an arbitrary ``well behaved'' test function (vanishing at the ends, sufficiently smooth). Expanding:
	\begin{equation}
		T(\epsilon)=\int_{x_1}^{x_2} F(x,h_0 + \epsilon\eta, h'_0 + \epsilon\eta') dx.
	\end{equation}
	At $h_0$, $\epsilon=0$, and $T'(0)=0$ for any $\eta$, as it is by definition the function that minimizes T.
	\begin{align}
		T'(\epsilon)&=\int_{x_1}^{x_2} \frac{d}{d\epsilon}F(x,h_0 + \epsilon\eta, h'_0 + \epsilon\eta') dx,\\
		&=\int_{x_1}^{x_2} \frac{\partial F}{\partial h}\eta + \frac{\partial F}{\partial h'}\eta' dx,\\
		&=\int_{x_1}^{x_2} \left(\frac{\partial F}{\partial h} +\frac{d}{dx} \frac{\partial F}{\partial h'}\right)\eta\ dx,
	\end{align}
	where the last equality comes from integration by parts on the second term, using the fact $\eta$ vanishes at the end points. (This is similar to the weak formulation for FEM and the continuous adjoint approach.) When $\epsilon = 0$,
	\begin{equation}
	T'(0) = \int_{x_1}^{x_2} \left(\frac{\partial F}{\partial h}(x, h_0, h_0') +\frac{d}{dx} \frac{\partial F}{\partial h'}(x, h_0, h_0') \right)\eta\ dx = 0.
	\end{equation}
	Because $\eta$ is arbitrary, for this integral to be zero, the bracketed term must be zero. In our case,
	\begin{align}
		F &= \frac{-\theta}{kh'},\\
		\frac{\partial F}{\partial h} &= 0.\\
		\frac{\partial F}{\partial h'} &= \frac{\theta}{kh'^2}.	
	\end{align} 
	Thus, it follows that
	\begin{equation}
	\boxed{k(x)h'^2(x) = \lambda \theta(x)\ \forall x}
	\label{eq: scaling min vel}
	\end{equation}
	where $\lambda$ is a constant. 	Head gradient, $h'(x)$ represents the local change in fluid energy density. As fluid flows downgradient, from higher to lower head, the porous media performs work on the fluid, dissipating its energy into heat via friction. The local \textit{rate} of this dissipation is then proportional to $q(x)h'(x)$: a higher flow rate means a larger volume of fluid has its energy reduced per unit time. $q(x)h'(x) = k(x)h'^2(x)$, which we found to have (assuming constant porosity) constant value $\lambda\theta$ under minimum travel time conditions. Thus, the minimum travel time head configuration corresponds to a constant rate of energy dissipation. 
	
\subsection{An explicit formula for minimum travel time}
	Rearranging \eqref{eq: scaling min vel} , knowing $h$ is a decreasing function,
	\begin{equation}
		h'(x) = -\lambda^{1/2}\theta^{1/2}(x) k^{-1/2}(x).
		\label{eq: h' from k}
	\end{equation}
	Substituting this into \eqref{eq: T}, we find
	\begin{equation}
	T=\lambda^{-1/2}\int_{x_1}^{x_2}\theta^{1/2}(x) k^{-1/2}(x) dx.
	\label{eq: T ito lambda}
	\end{equation}
	Integrating \eqref{eq: h' from k} between $x_1$ and $x_2$, we may solve for $\lambda^{1/2}$:
	\begin{equation}
		\lambda^{1/2} =  \frac{h(x_0)-h(x_1)}{\int_{x_1}^{x_2} \theta^{1/2}(x)k^{-1/2}(x) dx}
		\label{eq: lambda}
	\end{equation}
	Finally, combining \eqref{eq: T ito lambda} and \eqref{eq: lambda} yields an explicit expression for minimum travel time in terms of explicitly known functions:
	\begin{equation}
		\boxed{T=\frac{\left(\int_{x_1}^{x_2}\sqrt{\frac{\theta(x)}{k(x)}}dx\right)^2}{h(x_1)-h(x_2)} }
	\end{equation}

	\section{Head distribution for minimum energy dissipation in porous media}
	\label{sec: hd med}
	In general, potential flow configures itself to experience the least total resistance (energy dissipation) as it travels down gradient in accordance with Darcy's law, rather than least travel time, as in optics. It is then natural to consider the steady-state head distribution along a path that yields the least energy dissipation along that path, with the caveat that, in higher dimensions, energy is minimized globally, and not simultaneously on each path individually. Informed by the discussion above, we seek to select the function $h(x)$ to minimize \eqref{eq: E}
	Exactly following the approach in Section \ref{sec: cv mtt}, we seek to satisfy
	\begin{equation}
	\frac{\partial F}{\partial h} +\frac{d}{dx} \frac{\partial F}{\partial h'}= 0.
	\label{eq: euler equation}
	\end{equation}
	For \eqref{eq: E},
	\begin{align}
		F &= kh'^2,\\
		\frac{\partial F}{\partial h} &= 0,\\
		\frac{\partial F}{\partial h'} &= 2kh'.	
	\end{align} 
	Plugging in to \eqref{eq: euler equation} and integrating with respect to $x$, it follows that
	\begin{equation}
	\boxed{k(x)h'(x) = \lambda_E\ \forall x,}
	\label{eq: scaling min diss}
	\end{equation}
	where $\lambda_E$ is a constant.
	
	From Darcy's law, $q(x) \equiv -k(x)h'(x)$. It thus follows immediately from \eqref{eq: scaling min diss} that the head distribution that minimizes energy dissipation along the path from $x_1$ to $x_2$ results in a spatially constant Darcy velocity. Because groundwater flow is incompressible, stream tube aperture (and therefore velocity) responds to local fluctuations in hydraulic conductivity, this implies that a locally homogeneous conductivity in the vicinity of the streamline minimizes energy dissipation.

\section{Conductivity distribution for minimum travel time in porous media}
\label{sec: k mtt}
We directly consider the relation between conductivity variability and travel time by variational methods. We consider the case where we $h(x)$ is the natural head distribution resulting from the minimum dissipation principle, the head boundary conditions at $x_1$ and $x_2$, and the conductivity distribution $k(x)$, as solved in the last section. We here consider $k(x)$ as the function to be optimized, asking what $k(x)$ minimizes travel time, subject to the constraint that its mean value is fixed to some constant $\bar{k}$.

From \eqref{eq: scaling min diss}, we know that $h'(x) = \frac{q}{k(x)}$ It then follows:
\begin{align}
	h(x_2) - h(x_1) &= \int_{x_1}^{x_2} h'(x)\ dx\nonumber\\
	&= q\int_{x_1}^{x_2} \frac{dx}{k(x)}
\end{align}  
As Darcy velocity is constant, minimum travel time is achieved by minimizing
\begin{equation}
\frac{\theta}{q}=\frac{\int_{x_1}^{x_2} \frac{dx}{k(x)}}{h(x_2) - h(x_1)}.
\end{equation}
This is plainly a functional of $k(x)$ that can be minimized by calculus of variations. We define
\begin{equation}
	F(k) \equiv \frac{\theta(x)}{k(x)}
\end{equation}
However, we also need to constrain the mean value of $k(x)$ to $\bar{k}$. We define
\begin{equation}
	G(k) \equiv \int_{x_1}^{x_2}k\ dx - \bar{k}(x_2-x_1),
\end{equation}
so that the constraint is satisfied when $G=0$. We then define a Lagrangian
\begin{align}
	H(k,\gamma) &= F(k) + \gamma G(k),\nonumber\\
	&= \frac{1}{k(x)} + \gamma\left[k - \bar{k}(x_2-x_1)\right]
\end{align}
where $\gamma$ is an unknown Lagrange multiplier. A necessary condition for the constrained minimum to be achieved is satisfaction of the equations
\begin{eqnarray}
	\frac{\partial H}{\partial k} +\frac{d}{dx} \frac{\partial H}{\partial k'}&=& 0,\label{eq: augmented euler}\\
	\frac{\partial H}{\partial \gamma} &=& 0.
\end{eqnarray} 
It follows immediately from \eqref{eq: augmented euler} that
\begin{equation}
	\frac{1}{k^2} +\frac{1}{k}\frac{\partial \theta}{\partial k} = \gamma.
	\label{eq: k relation}
\end{equation}
Assuming that porosity is spatially constant, or that it relates to conductivity in Kozeny--Carman fashion \citep[$\theta \sim\propto k^{1/3}$][]{Pape2000}, the LHS of \eqref{eq: k relation} is a decreasing function of $k$, for $k>0$. Thus \eqref{eq: k relation} has a unique solution and the hydraulic conductivity must be constant for the minimum travel time to be achieved:
\begin{equation}
\boxed{k(x)=\bar{k}\ \forall x.}
\end{equation}
This aligns with the analysis of \cite{Kacimov2009}.

\section{Minimum energy dissipation in hydraulically parallel fractures}
\label{sec: md fracture}
\begin{figure}
	\centering
	\begin{tikzpicture}
		\draw [line width=1mm, black,  decorate, decoration={random steps}] (-1,3) -- node [below, font=\fontfamily{phv}\selectfont] {Pathway A} (9,3.1);
		\draw [line width=1.5mm, black, decorate, decoration={random steps}] (-1,1.2) -- node [below, font=\fontfamily{phv}\selectfont] {Pathway B} (9,1);
		\draw [line width=1mm, gray] (0,0) -- (0.5,4);
		\draw [line width=1mm, gray] (8,0) -- (8.5,4);
		\draw [fill=green] (0.4,3.05) circle [radius=1.5mm] node [below right] {$x_1$};
		\draw [fill=red] (8.1,1) circle [radius=1.5mm] node [below right] {$x_2$};
		\draw[->, line width=1mm, blue] (3,2) -- (4,2) node [right, font=\fontfamily{phv}\selectfont] {Flow direction};
	\end{tikzpicture}
	\caption{Schematic of flow from node $x_1$ to node $x_2$, where these are connected by two hydraulically parallel flow pathways, A and B. The pathways here consist of quasi-horizontal fractures (black lines) intersected by quasi-vertical joints (gray lines).}
	\label{fig: parallel frac}
\end{figure}
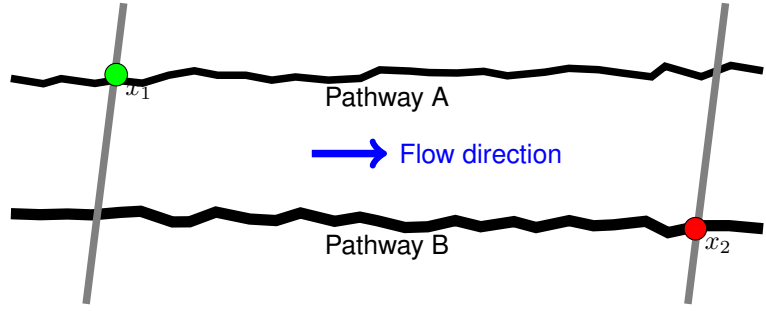
	\cite{Narasimhan1999} showed that the solution to the steady-state groundwater flow equation, $\nabla\cdot (k\nabla \mathbf{q})=0$, in an arbitrary domain minimized energy dissipation in that system, subject to a given set of boundary conditions. However, his argument was inherently restricted to a porous medium continuum. We here establish an equivalent result for fracture networks consisting of multiple parallel paths. Because multiple parallel pathways can be mathematically consolidated into single pathways with an effective transmissivity, we without loss of generality consider a two-pathway system.

	Consider two hydraulically parallel fracture pathways, $A$ and $B$, with respective average transmissivities $\tau_A$ and $\tau_B\ \mathrm{[L^2T^{-1}]}$, joining nodes $x_1$ and $x_2$ whose head difference is $\Delta h\ \mathrm{[L]}$. An example configuration is shown in Figure \ref{fig: parallel frac}. Let $Q_B$ and $Q_B\ \mathrm{[L^3T^{-1}]}$ be the flow rates in these fracture pathways. Both pathways share the same head drop, and assuming flow in both is Darcian, it follows that 
	\begin{equation}
		\frac{Q_A}{Q_B} = \frac{\tau_A}{\tau_B}.
		\label{eq: K Q ratio}
	\end{equation} 
	Recall that head is proportional to energy per unit volume, with constant of proportionality $\rho g$. Thus, energy dissipated per unit time, $E_d\ \mathrm{[ML^2T^{-3}]}$, for a fracture system with head drop $\Delta h$ satisfies
	\begin{equation}
		\frac{E_d}{\rho g} = Q\Delta h.
	\end{equation}
	So minimizing energy dissipation in the two fractures is equivalent to minimizing
	\begin{equation}
		\frac{E_{d,\mathrm{sys}}}{\rho g} \equiv \left(Q_A+Q_B\right)\Delta h.
	\end{equation}

	Consider a perturbation $\pm\Delta Q$ in the flow partitioning between the fractures that preserves total flow rate, so that
	\begin{eqnarray}
		Q_A &\rightarrow& Q_A + \Delta Q\\
		Q_B &\rightarrow& Q_B - \Delta Q,
	\end{eqnarray}
	Assuming the cubic law continues to apply in each fracture pathway individually, the total energy dissipation may be assessed by assigning pathway A fictitious head drop $\Delta h + \frac{\Delta Q}{\tau_A}$ and pathway B fictitious head drop $\Delta h - \frac{\Delta Q}{\tau_B}$. This is to say: we imagine that the parallel pathways are disconnected at node $x_1$, and an independent head drop is assigned to each pathway to achieve the desired flow rate. The rate of energy dissipated by the perturbed system as a whole, $E^*_{d,\mathrm{sys}}$, satisfies
	\begin{eqnarray}
		\frac{E^*_{d,\mathrm{sys}}}{\rho g} &=& \left(Q_1 + \Delta Q\right)\left(\Delta h + \frac{\Delta Q}{\tau_A}\right) + \left(Q_B - \Delta Q\right)\left(\Delta h - \frac{\Delta Q}{\tau_B}\right)\\
		&=& \left(Q_A+Q_B\right)\Delta h+\Delta Q\cancelto{0}{\left(\Delta h-\Delta h\right)}+\Delta Q \cancelto{0}{\left(\frac{Q_A}{\tau_A}-\frac{Q_B}{\tau_B}\right)}+(\Delta Q)^2\left(\frac{1}{\tau_A}+\frac{1}{\tau_B}\right) \label{eq: refactor}\\
		&=&\frac{E_{d,\mathrm{sys}}}{\rho g}+(\Delta Q)^2c\left(\frac{1}{\tau_A}+\frac{1}{\tau_B}\right),\label{eq: energy final}
	\end{eqnarray}
	where the second cancellation in \eqref{eq: refactor} follows from \eqref{eq: K Q ratio}. Because the second term of \eqref{eq: energy final} is always positive, it follows that $E^*_{d,\mathrm{sys}} \ge E_{d,\mathrm{sys}}$. Thus the natural flow partitioning minimizes energy dissipation.

\section{Minimum travel time in sequential fractures}
\label{sec: cubic}
\subsection{Formulation}
	Consider a path through a fracture network consisting of $N$ segments of length $\Delta x\ \mathrm{[L]}$ in the direction of flow, all with identical width transverse to flow, $w\ \mathrm{[L]}$. (This model is general enough to accommodate variable-aperture fractures discretized to resolution $\Delta x$.) By Darcy's law and the cubic law, at each point in the fracture system
	\begin{equation}
		Q=c w a_i^3\frac{\Delta h_i}{\Delta x},
		\label{eq: Q}
	\end{equation}
	where $Q\ \mathrm{[L^3T^{-1}]}$ is the volumetric flow rate, $c\ \mathrm{[L^{-1}T^{-1}]}$ is a constant, $a_i\ \mathrm{[L]}$ is the fracture aperture in segment $i$, $\Delta h_i\ \mathrm{[L]}$ is the head drop across segment $i$, 
	By continuity---fluid is assumed incompressible---$Q$ is constant throughout all segments. We also assume that there is a fixed head drop, $\Delta H\ \mathrm{[L]}$, across the whole flow path.
	
	The travel time, $T\ \mathrm{[T]}$, through the system is the sum of the travel times through each segment:
	\begin{equation}
		T=\sum_{i=1}^N \frac{w a_i \Delta x}{Q(a_1,\dots a_N)}.
		\label{eq: T frac}
	\end{equation}
	$Q$ is implicitly a function of all the apertures, because these collectively determine the effective conductivity of the network. If the aperture is increased in any of the segments, regardless of the current configuration, the effective hydraulic conductivity of the network path will increase, and so will $Q$. However, for a given value of $Q$ for the system, increased aperture in one segment will \textit{decrease} the effective velocity in said segment. We are interested in (a) understanding when one effect is stronger than the other, so that a perturbation results in a decreased travel time through the network, particularly when starting from a spatially uniform aperture and (b) given an average aperture, finding the distribution of apertures that will result in minimum advection time.
	
\subsection{Travel time change in response to local perturbation}
	Differentiate the total travel time with respect to the aperture, $a_n$ of arbitrary segment $n$, applying the product rule:
	\begin{equation}
		\frac{\partial T}{\partial a_n} = \frac{w \Delta x}{Q} - \frac{w \Delta x \sum a_i}{Q^2}\frac{\partial Q}{\partial a_n}.
		\label{eq: dTdan}
	\end{equation}
	
\begin{figure}
		\centering
		\includegraphics[width=0.75\textwidth]{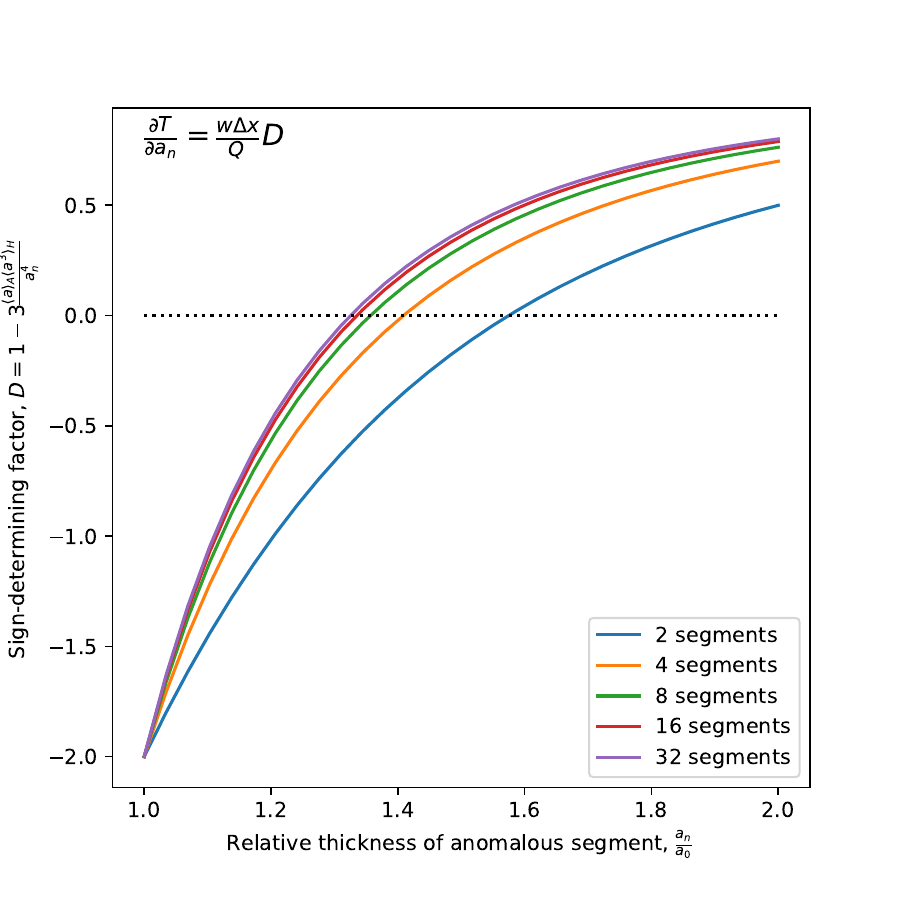}
		\caption{Change in travel time as a function of relative width of single anomalous segment. For small increases in relative width, $D$ is negative (below dotted line), meaning that travel time decreases with increasing relative width.}
		\label{fig: sign}
	\end{figure}

	To determine $\frac{\partial Q}{\partial a_n}$ we must write $Q$ as an explicit function of the various $a_i$. Rearranging \eqref{eq: Q}, it follows that for all segments:
	\begin{equation}
		\Delta h_i = \frac{Q\Delta x}{cw}\frac{1}{a_i^3}.
	\end{equation}
	Summing over all segments
	\begin{equation}
		\Delta H = \frac{Q\Delta x}{cw}\sum_{i=1}^N \frac{1}{a_i^3}
	\end{equation}
	Defining the total fracture / network length $L\equiv N \Delta X$:
	\begin{equation}
		\frac{\Delta H}{L} = \frac{Q}{cw} \left[\frac{1}{N}\sum_{i=1}^N \frac{1}{a_i^3}\right].
	\end{equation}
	\begin{eqnarray}
		Q &=& cw\left[\frac{1}{N}\sum_{i=1}^N \frac{1}{a_i^3}\right]^{-1}\frac{\Delta H}{L}\\
		&=&  cw\left<a^3\right>_H\frac{\Delta H}{L},
		\label{eq: Q ito a}
	\end{eqnarray}
	where $\left<\cdot\right>_H$ denotes the harmonic mean. Thus, we can compute
	\begin{eqnarray}
		\frac{\partial Q}{\partial a_n} &=& -cw\left[\frac{1}{N}\sum_{i=1}^N \frac{1}{a_i^3}\right]^{-2}\frac{\Delta H}{L}\left[\frac{-3}{N a_n^4}\right]\\
		&=& cw\left<a^3\right>_H^2\left[\frac{3}{N a_n^4}\right]\frac{\Delta H}{L},\\
		&=& Q\left<a^3\right>_H\left[\frac{3}{N a_n^4}\right],
		\label{eq: dqda ito a}
	\end{eqnarray}
	where the last equality follows from \eqref{eq: Q ito a}. By substituting \eqref{eq: dqda ito a} into \eqref{eq: dTdan}, we see 
	\begin{eqnarray}
		\frac{\partial T}{\partial a_n} &=& \frac{w \Delta x}{Q} - \frac{w \Delta x \sum a_i}{Q^2}Q\left<a^3\right>_H\left[\frac{3}{N a_n^4}\right]\\
		&=& \frac{w \Delta x}{Q} \left(1 - 3\frac{\left<a\right>_A\left<a^3\right>_H}{a_n^4}\right),
		\label{eq: final sensitivity}
	\end{eqnarray}
	where $\left<\cdot\right>_A$ denotes the arithmetic mean. In a homogeneous fracture or system, $a_n=\left<a\right>_A$ and $a_n^3=\left<a^3\right>_H$. Thus, when $a_n$ deviates from a homogeneous aperture,
	\begin{equation}
		\boxed{\frac{\partial T}{\partial a_n} = \frac{-2 w \Delta x}{Q}.}
	\end{equation}
	Thus, for all uniform-aperture, fixed-width fractures or sequences of fractures obeying the cubic law, a small increase in the aperture of one portion will decrease the total travel time through the whole fracture or fracture sequence. Note that this is not necessarily true for heterogeneous fractures. If $a_n \gg \left<a\right>_A$, $\left<a^3\right>_H$ is largely independent of perturbations in $a_n$, and the RHS of \eqref{eq: final sensitivity} may be positive. In Figure \ref{fig: sign}, we see the change in travel time as a function of aperture of a single anomalous segment.
	
\subsection{Aperture distribution for minimum travel time}
	We consider the minimum advective travel time subject to the constraint $\left<a\right>_A=\bar{a}$, where $\bar{a}\ \mathrm{[L]}$ is a fixed mean aperture. We formulate the Lagrangian by augmenting $T$, as defined in \eqref{eq: T frac} with the constraint:
	\begin{equation}
`		L \equiv T(a_1,\dots,a_N) + \lambda \left(\sum_{i=1}^N a_i - N\bar{a}\right),
	\end{equation}
	where $\lambda$ is a Lagrange multiplier. A necessary condition for an extreme value to be obtained subject to the constraint is that, for all $n$,
	\begin{eqnarray}
		0 &=& \frac{\partial L}{\partial a_n}\\
		&=& \frac{\partial T}{\partial a_n} + \lambda\frac{\partial}{\partial a_n}\left(\sum_{i=1}^N a_i - N\bar{a}\right)\\
		&=&  \frac{w \Delta x}{Q} \left(1 - 3\frac{\left<a\right>_A\left<a^3\right>_H}{a_n^4}\right) + \lambda.
	\end{eqnarray} 
	Thus for all $n$,
	\begin{equation}
		a_n^4 = \left(\frac{\lambda Q}{w\Delta x} + 1\right)
	\end{equation}
	This has a single real solution, so the unique extremum is found to exist when $a_n = \bar{a}\ \forall n$. It is obvious that this extreme value is a minimum because $T$ can be made arbitrarily large, subject to the mean constraint by setting $a_0=0$ and altering the other apertures accordingly.
	
	Thus it is simultaneously true that the shortest advection time for a given mean aperture is in a uniform-aperture system, \textit{and} that any sufficiently small positive local aperture perturbation from a uniform-aperture system will decrease advection time.  

\section{Summary and conclusions}
	We examined advective travel time and energy dissipation for groundwater flow in heterogeneous hydraulic permeability fields and fracture systems, identifying conditions under which both are minimized. Steady-state flow under natural gradient in both types of systems satisfy a straightforward energy minimization condition. Minimum energy dissipation along a streamline occurs under conditions in which flow velocity is constant along the streamline (implying local homogeneity of conductivity or transmissivity), and under fixed boundary conditions and mean conductivity, a homogeneous conductivity field was shown to globally minimize energy dissipation. Where a streamline passes through an arbitrary head field, optimal advective travel time is found to occur under conditions of constant energy dissipation along the streamline. In hydraulically parallel fractured systems obeying the cubic law, we showed that the standard Darcy flow solution minimized energy dissipation among all possible flow partitions. In sequential fracture systems, we showed found that a uniform aperture system minimizes advective transport time from among all systems with the same length and average aperture. However, we also found that any sufficiently small positive local perturbation about a uniform aperture reduces advection times further. Besides their basic scientific interest, these results may have relevance for design of pumping interventions into groundwater flow systems to improve transport rates or energy losses.

\bibliographystyle{apalike}
\bibliography{mvmd}

\end{document}